# Large Effects of Electric Fields on Atom-Molecule Collisions at Millikelvin Temperatures

L. P. Parazzoli[1], N. J. Fitch[1], P. S. Żuchowski[2], J. M. Hutson[2], H. J. Lewandowski[1]

[1] *JILA and Department of Physics, University of Colorado, Boulder, Colorado 80309-0440*
[2] *Department of Chemistry, University of Durham, Durham, DH1 3LE, UK*
(Dated: January 14, 2011)

Controlling interactions between cold molecules using external fields can elucidate the role of quantum mechanics in molecular collisions. We create a new experimental platform in which ultracold rubidium atoms and cold ammonia molecules are separately trapped by magnetic and electric fields and then combined to study collisions. We observe inelastic processes that are faster than expected from earlier field-free calculations. We use quantum scattering calculations to show that electric fields can have a major effect on collision outcomes, even in the absence of dipole-dipole interactions.

At temperatures below 1 kelvin, molecular collisions enter a new quantum regime governed by long-range interactions and threshold effects. Studying collisions in this regime offers exciting possibilities for controlling chemical processes. For example, the electric dipole moment of cold polar molecules can serve as a handle to orient the molecules in space and allows us to explore the stereodynamics of chemical reactions [1–4]. Recent ground-breaking experiments using trapped KRb molecules at 250 nK have shown how quantum statistics and dipole-dipole interactions can affect bimolecular reaction rates in a fully quantum-mechanical regime [5, 6]. However, these experiments deal with interactions at very long range, while for chemical reactions, one would like to understand and control interactions at short range where chemical forces are relevant and one can hope to influence reaction pathways. Such experiments would be valuable for a wide range of molecules, but unfortunately the technology used to produce molecular gases below 1 $\mu$K is currently applicable only to diatomic alkali molecules.

Crossed molecular-beam experiments, by contrast, can be carried out with a large variety of chemically important molecules. Most of what we know about the collision dynamics of small molecules in the gas phase comes from such experiments [7–9]. They have allowed detailed studies of reaction mechanisms involving both direct scattering and complex formation [7], identified reactive-scattering resonances [8], and probed the time dependence of chemical reactions [9]. However, the collision energy in crossed-beam experiments is usually greater than 100 K, though some experiments at lower energies have been carried out using guided beams [10–13], Laval nozzles [14], and traps [15, 16]. Experiments at these higher temperatures do not probe the quantum regime where detailed control is possible.

In this letter we introduce a versatile platform for investigating atom-molecule interactions at temperatures below 1 K and demonstrate, with the aid of theory, that an electric field can strongly affect cold atom-molecule collisions. Our approach incorporates the advantages of molecular beams for cooling a wide variety of chemically important molecules and the benefits of using ultracold atoms to reach low collision energies where only a few partial waves contribute to collisions. We produce cold trapped molecules using a slowed molecular beam and an electrostatic trap. In a separate part of the same vacuum chamber, we use laser cooling and magnetic trapping to produce an ultracold gas of atoms. The two traps are then spatially overlapped and collisions are monitored through measurements of trap loss. As a demonstration of this combined ultracold atom-cold molecule approach, we study collisions of deuterated ammonia (ND$_3$) with rubidium (Rb) and explore a competition between the orienting effect of the intermolecular forces and those of the electric field.

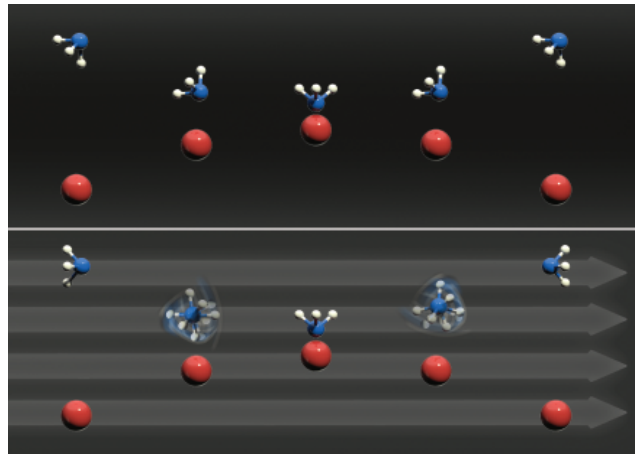

FIG. 1: Diagram of the collisions without (top) and with (bottom) an electric field. At zero field, ND$_3$ orients smoothly along the intermolecular axis, and there is a low probability of reorientation after a collision with Rb. However, when an electric field (grey arrows) is applied, the orientation of the ND$_3$ becomes confused when the potential anisotropy is similar to the electric Stark energy, leading to an increased probability of a state-changing collision.

Ammonia is an important molecule in both planetary atmospheres and in the interstellar medium [17, 18]. In its rovibrational ground state, it has a trigonal pyramidal structure. In the absence of electric fields, the two lowest-energy eigenstates of para-ND$_3$ are symmetric and antisymmetric combinations of two equivalent umbrella states, in which the nitrogen is either above or below the plane defined by the hydrogen atoms. Because there is a barrier for the nitrogen to tunnel through this plane, the two eigenstates are separated in energy by the tunneling splitting (0.053 cm$^{-1}$), with the symmetric combination being lower in energy. The eigenstates are denoted by $|jkm\pm\rangle$, where $j$ is the rotational angular momentum, $k$ and $m$ are the projections of $j$ onto the symmetry axis of the molecule and onto the laboratory field axis, and $\pm$ labels the parity-adapted states of the tunneling doublet. An electric field mixes these two states, causing the energy separation between the states to increase. In our experiment, molecules in the low-field seeking $|111u\rangle$ upper state are trapped in static electric fields. However, the lower state $|111l\rangle$ is high-field seeking and cannot be trapped with static fields. Inelastic collisions that transfer molecules from the upper to the lower state therefore result in loss of molecules from the trap.

Scattering processes at low temperatures are sensitive to the details of intermolecular potential energy surfaces. The surface for Rb-ND$_3$ is highly anisotropic, with a potential well 1800 cm$^{-1}$ deep when the Rb approaches the ND$_3$ from the nitrogen side [19]. The large anisotropy causes strong coupling between the ND$_3$ rotation-tunneling states during a collision. Because of this, we initially expected that the inelastic cross sections would be well approximated by the semiclassical Langevin limit, which assumes that any collision that crosses the centrifugal barrier leads to an inelastic scattering event [20]. The Langevin-limited cross section at 100 mK is approximately 4000 Å$^2$. However, quantum-mechanical calculations on Rb-ND$_3$ collisions in the absence of electric fields [21] gave inelastic cross sections for de-excitation from the upper to the lower component of the tunneling doublet about a factor of 10 below this value. The suppression arises because, in the absence of an electric field, the first effect of the potential anisotropy is to mix the components of the tunneling doublet to form umbrella states quantized along the intermolecular axis, in which the ND$_3$ pyramid points either towards or away from the Rb atom (Fig. 1, top). The two resulting states are well separated in energy at all intermolecular distances, so that collisions occur almost adiabatically. Inelastic transitions between the two tunneling states are quite weak, with cross sections on the order of a few hundred square Ångstroms at 100 mK [21].

The collision dynamics change greatly when an electric field is applied (Fig. 1, bottom). In this case, there are two competing forces. At long range (> 30 bohr), the electric field mixes the two tunneling states and orients the ND$_3$ either with or against the electric field direction, regardless of the direction of the incoming Rb. Conversely, at short range (6-15 bohr), the anisotropy of the interaction potential forces the ND$_3$ to align along the intermolecular axis. At some intermediate value of the intermolecular distance, the Stark energy is comparable to the potential anisotropy, and there is a competition between the two forces. The competition between aligning with the field and aligning with the collision axis in this region allows nonadiabatic transitions between states. Through this process, the externally applied electric field increases the inelastic collision cross section. We present quantum-mechanical calculations below demonstrating this effect.

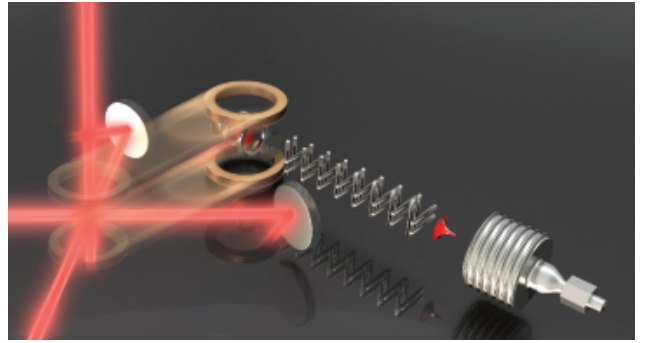

FIG. 2: Schematic representation of the experimental apparatus (not to scale). The atoms are cooled and trapped at the intersection of laser-cooling beams. A beam of cold molecules is created by a pulsed valve (lower right). The molecular beam is slowed and trapped using electric fields created by ∼150 in-vacuum electrodes (metallic rods and rings). The coils forming the atomic trap are then translated across the table to overlap the atomic and molecular traps.

We experimentally determine both the elastic and inelastic cross sections between $^{14}$ND$_3$ and $^{87}$Rb. At collision energies of 100 mK, the only accessible inelastic scattering pathways are $|111u\rangle \to |11ml\rangle$ and $|111u\rangle \to |110u\rangle$, which are all exothermic. The states created by these inelastic processes are not low-field seeking, and the molecules are therefore ejected from the trap. At the same time that inelastic collisions are causing trap loss, elastic collisions between the ND$_3$ and the (much colder) Rb atoms result in a cooling of the trapped ND$_3$. This cooling increases the ND$_3$ density at the trap center. Measurements of the time dependence of the ND$_3$ density at the trap center are thus able to probe both the elastic and inelastic collisions.

The cold ND$_3$ and ultracold Rb are initially prepared in two distinct regions of the vacuum system and then combined for the interactions to take place (Fig. 2). First, the ultracold atom sample is prepared by standard laser cooling using a magneto-optic trap. The cold atoms are then transferred to a quadrupole magnetic trap. Second, we prepare a sample of cold molecules by Stark deceleration [22], in which inhomogeneous time-varying electric fields are used to decelerate a supersonic beam of $ND_3$ molecules seeded in a carrier gas of krypton. The molecules are decelerated from a mean speed of 415 m/s to 32 m/s before they enter a region of the vacuum system containing high-voltage electrodes that form a quadrupole electrostatic trap. During the trap-loading process, the remaining translational energy is removed, producing a trapped sample of $ND_3$ in the $|111u\rangle$ state. Finally, to carry out the collision experiment, the coils that trap the atoms are translated across the optical table to overlap the atomic and molecular clouds [23].

The properties of the trapped molecular gas are determined by resonantly ionizing the $ND_3$ and detecting the resulting ions using a microchannel plate detector. The measured lifetime of the $ND_3$ in the trap is around 1 s and is limited by collisions with background gas atoms at 300 K (Fig. 3A). In addition to determining the time-dependence of the number density, the ionization laser can be scanned along one axis of the trap to measure the density distribution, allowing us to extract a temperature of the gas. While not fully in thermal equilibrium, the distribution of the trapped $ND_3$ can be characterized by a temperature of 100 mK, with a peak density of $10^6$ molecules/cm$^3$. The atomic gas is probed with resonant light to determine its temperature (600 $\mu$K) and peak density ($10^{10}$ cm$^{-3}$). Both gas clouds occupy approximately the same volume of 1 mm$^3$.

Once we have both $ND_3$ and Rb trapped and co-located, we study their collisions through population decay. Fig. 3A shows the decay of the $ND_3$ central density for a variety of initial Rb densities. The Rb-enhanced decay has a complex form because of the time dependence of the Rb number and temperature, inhomogeneity of the density of both gases, and the role of momentum-changing elastic collisions. We model this with Monte Carlo simulations in which the Rb density is modeled as a mean field whose time-dependent phase-space distribution is derived from the measured temperature and peak density, while 50,000 $ND_3$ trajectories are individually propagated in time. Probabilities for inelastic or elastic scattering events are calculated for each time step based on the molecular velocity, collision cross sections, and local Rb density. We run 250 simulations varying the elastic and inelastic cross sections for each experimental measurement. The reduced chi-squared from the resulting simulated decay curves is used to determine the fitness of each simulation. Fig. 3C shows the collision cross sections obtained from our measurements.

Our measurements allow us to place an upper limit on the elastic cross section. Elastic collisions with the much colder Rb reduce the average energy of the $ND_3$ gas and consequently increase the density of $ND_3$ at the center of the trap. In the case of Rb-$ND_3$, the elastic and inelastic collision rates are similar, so that the temperature of the $ND_3$ does not change substantially. Nevertheless, the small change in the spatial distribution of $ND_3$ plays a measureable role in the trap dynamics. In Fig. 3B, we plot the results of simulations with varying elastic cross sections with a fixed value of 2000 Å$^2$ for the inelastic cross section. The effect on the decay profile due to elastic collisions with Rb may be easily seen. The ability to distinguish elastic and inelastic collisions is one of the major advantages of the double-trap system.

Our measured inelastic cross section of 2000 Å$^2$ is much larger than the field-free calculations [21] and only about a factor of two below the Langevin limit. To explain this, we have developed a new computer code to carry out quantum-mechanical scattering calculations on Rb-$ND_3$ collisions in the presence of an electric field. The code performs coupled-channel calculations in a fully decoupled basis set of space-quantized functions $|jkm\pm\rangle|LM_L\rangle$, where $L$ is the angular momentum for rotation of Rb and $ND_3$ about one another and $M_L$ is the projection of $L$ onto the electric-field axis. The electron spin of the Rb atom and all the nuclear spins are neglected. The electric field destroys the quantum numbers for the total parity, $p$, and the total angular momentum, $J$, so that all basis functions for a particular value of $M = m + M_L$ must be included simultaneously. This requires the solution of very large sets of coupled equations and makes fully converged calculations impractical. Nevertheless, calculations with a reduced basis set ($j_{\max} = 4, k = \pm 1, L_{\max} = 11$) are sufficient to illustrate the qualitative effects. The resulting elastic and total inelastic cross sections from low-field-seeking states of $ND_3$ are shown in Fig. 4A; the effect of an electric field is to increase the inelastic cross sections substantially, while also decreasing the elastic cross sections.

The large increase above the field-free prediction comes from two sources. One of these is trivial, and arises from the fact that collisions that change $m$ without altering other $ND_3$ quantum numbers are elastic at zero field but inelastic when a field is applied. However, the major contribution is a genuine dynamical effect that arises from the applied field. In the absence of a field, the $J$ and $p$ quantum numbers are conserved. Fig. 4B shows potential energy curves obtained by diagonalizing the rotation-tunneling Hamiltonian as a function of the intermolecular distance $R$, with the tunneling splitting set to zero for simplicity. The curves for different $J$ values are color-coded and cross one another with no mixing. However, when an electric field is introduced (Fig. 4C), the $ND_3$ levels with different values of $m$ become nondegenerate, and the curves corresponding to trapped and untrapped states display an extensive network of avoided crossings. In dynamical terms, this effect may be seen as arising from competition between the potential anisotropy, which tries to quantize the $ND_3$ tunneling along the intermolecular axis, and the applied field, which tries to quantize it along the space-fixed field axis. As the Rb and $ND_3$ approach one another, there is always a point where the two effects balance, and the axis of quantization is indeterminate.

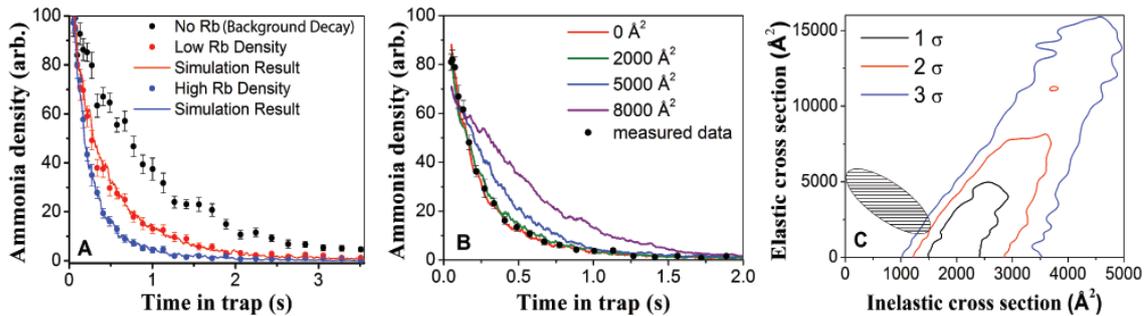

FIG. 3: Measured decay curves and simulation results for $ND_3$ and extracted elastic and inelastic cross sections. (A) The peak ammonia density is plotted as a function of time for various densities of Rb, including the corresponding best-fit simulations. (B) The effect of elastic collisions on the simulated $ND_3$ decay profile. (C) Confidence contours for the elastic and inelastic cross sections extracted from simulations of the experimental decay curves. The shaded oval represents the range of possible results from field-free quantum-mechanical scattering calculations and is obtained by repeating the calculations for a variety of scaled potential-energy surfaces.

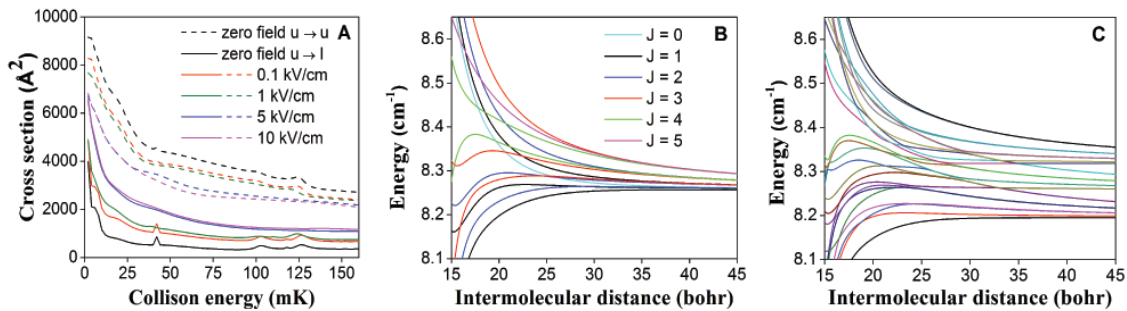

FIG. 4: (A) Elastic (dashed) and inelastic (solid) cross sections at a variety of electric fields, obtained from quantum-mechanical scattering calculations on the potential energy surface of Ref. [19]. Pure $m$-changing collisions are elastic at zero field, but inelastic at finite field. (B) Adiabatic potential curves for zero-field collisions, showing crossings between curves for different total angular momentum $J$. The isotropic average potential $V_{00}(R)$ is subtracted so that the curves stay within a narrow energy range. The zero-field curves are independent of $M$. (C) Same as (B), but in a field of 5 kV/cm, showing an extensive series of avoided crossings where reorientation can occur. Only curves for $M=1$ are shown.

The combination of experimental and theoretical results shows that, even when only one of the colliding species is polar, electric fields may have a major effect on collision dynamics at millikelvin temperatures. In the future, as experiments of molecular collisions and reactions push further into the cold and ultracold temperature regime, it will be possible to use electric fields to control collision processes.

This work has been supported by NSF, AFOSR, the Petroleum Research fund, the A. P. Sloan Foundation, and EPSRC under the ESF EuroQUAM Program.

[1] R. N. Zare, Science **279**, 1875 (1998).
[2] G. Quéméner and J. L. Bohn, Phys. Rev. A **81**, 022702 (2010).
[3] M. H. G. de Miranda et al., arXiv:1010.3731 (2010).
[4] R. V. Krems, Phys. Chem. Chem Phys. **10**, 4079 (2008).
[5] S. Ospelkaus et al., Science **327**, 853 (2010).
[6] K. K. Ni et al., Nature **464**, 1324 (2010).
[7] G. Scoles, *Atomic and Molecular Beam Methods*, vol. 1 (Oxford University Press, Oxford, 1988).
[8] R. T. Skodje et al., Phys. Rev. Lett. **85**, 1206 (2000).
[9] S. C. Althorpe et al., Nature **416**, 67 (2002).
[10] M. Kirste et al., Phys. Rev. A **82**, 042717 (2010).
[11] L. Scharfenberg et al., Phys. Chem. Chem. Phys. **12**, 10660 (2010).
[12] B. C. Sawyer et al., arXiv:1008.5127 (2010).
[13] J. J. Gilijamse et al., Science **313**, 1617 (2006).
[14] I. W. M. Smith, Angew. Chem. Int. Ed. **45**, 2842 (2006).
[15] M. T. Hummon et al., arXiv:1009.2513 (2010).
[16] E. R. Hudson et al., Phys. Rev. Lett. **100**, 203201 (2008).
[17] D. Sudarsky et al., Astrophys. J. **538**, 885 (2000).
[18] P. T. P. Ho and C. H. Townes, Annu. Rev. Astron. Astrophys. **21**, 239 (1983).
[19] P. S. Żuchowski and J. M. Hutson, Phys. Rev. A **78**, 022701 (2008).

[20] R. D. Levine, *Molecular Reaction Dynamics* (Cambridge University Press, Cambridge, 2005).
[21] P. S. Żuchowski and J. M. Hutson, Phys. Rev. A **79**, 062708 (2009).
[22] H. L. Bethlem, G. Berden, and G. Meijer, Phys. Rev. Lett. **83**, 1558 (1999).
[23] H. J. Lewandowski *et al.*, J. Low Temp. Phys. **132**, 309 (2003).